\documentclass[a4paper]{jpconf}
\usepackage{graphicx}
\usepackage{graphicx}
\usepackage{amssymb}

\newcommand{\Gyr}{{\,\rm Gyr}}
\newcommand{\pc}{\,\mathrm{pc}}

\newcommand{\mE}{\mathcal{E}}
\newcommand{\Mbh}{M_{\bullet}}

\newcommand{\Mo}{M_{\odot}}

\newcommand{\Ms}{M_{\star}}

\newcommand{\tp}{t_{\omega}}
\newcommand{\Jp}{J_{\omega}}
\newcommand{\apj}{ApJ}
\newcommand{\aj}{AJ}
\newcommand{\mnras}{MNRAS}

\begin{document}

\title{Resonant Relaxation near the Massive Black Hole in the Galactic Center}

\author{Clovis Hopman and Tal Alexander}

\address{Faculty of Physics, Weizmann
Institute of Science, POB 26, Rehovot 76100, Israel}

\ead{clovis.hopman@weizmann.ac.il}

\begin{abstract}
The coherent torques between stars on orbits near massive black holes
(MBHs) lead to resonant angular momentum relaxation. Due to the fact
that orbits are Keplerian to good approximation, the torques
efficiently change the magnitude of the angular momenta and rotate the
orbital inclinations. As a result the stars are rapidly
randomized. The galactic MBH is a good system for the observational
study of resonant relaxation. The age of the young B-stars at a
distance of $\sim0.01\pc$ from the MBH is comparable to the resonant
relaxation time, implying that resonant relaxation may have played an
important role in their dynamical structure. In contrast, the O-stars
in the stellar disks at $\sim0.1\pc$ are younger than the resonant
relaxation time, as required by their dynamical coherence. Resonant
relaxation dynamics dominates the event rate of gravitational wave
(GW) emission from inspiraling stars into MBHs of masses comparable to
the Galactic MBH. Resonant relaxation leads to rates $\lesssim\!10$
times higher than those predicted by 2-body relaxation, which would
improve the prospects of detecting these events by future GW detectors,
such as \textit{LISA.}

\end{abstract}

\section{Introduction}

In ``collisional'' stellar dynamics the potential in which a star
moves is considered to be smooth to first order, and the fact that the
potential is in fact made out of discrete stars is treated as a
perturbation (e.g. Chandrasekhar \cite{Ch43}; Binney \& Tremaine
\cite{BT87}). Stars have orbital parameters such as angular momentum
and energy that are conserved in the smooth potential, and these
parameters indeed remain constant for many dynamical times $t_d$. Only
the weak encounters allow for changes in these quantities. For
example, two stars which interact with each other, can exchange energy
and angular momentum, so that after their encounters their orbits are
described by slightly different quantities. In this way not only the
orbits of individual stars are modified, but the distribution function
(DF) of the entire system can change, and evolve towards a steady
state. The time-scale over which a system evolves is the {\it
relaxation time} $t_r$. In most systems $t_d\ll t_r$, and relaxation
can indeed be treated as a second order effect.

Analyses of the evolution of the DF near a MBH have almost exclusively
relied on the assumption that the mechanism through which stars
exchange angular momentum and energy is dominated by
\textit{uncorrelated two-body interactions}\footnote{$N$-body
simulations form an exception; for $N$-body simulations near MBHs see
e.g. Baumgardt, Makino \& Ebisuzaki \cite{Baum04a}, \cite{Baum04b};
Preto, Merritt \& Spurzem \cite{P04}; Merritt \& Szell
\cite{MS05}}. Any encounter is assumed to be unrelated to previous and
future encounters, and changes in energy and angular momenta are
considered to be drawn from a specified random
distribution. Relaxation can therefore in a meaningful way be
considered to be a random walk process. In the context of stellar
dynamics near MBHs, the assumption of uncorrelated encounters is made
in Fokker-Planck models (e.g. Bahcall \& Wolf \cite{BW76}; Bahcall \&
Wolf \cite{BW77}; Cohn \& Kulsrud \cite{CK78}; Murphy, Cohn \& Durisen
\cite{MCD91}), where the microscopic interactions are expressed by the
diffusion coefficients, and in Monte Carlo simulations (e.g. Shapiro
\& Marchant \cite{SM79}; Marchant \& Shapiro \cite{MS79}, \cite{MS80};
Freitag \& Benz \cite{FB01}, \cite{FB02}). Stars around MBHs are
described as moving in the smooth average potential of the MBH and the
stars, and the scattering by the fluctuating part of the potential is
modeled as a hyperbolic Keplerian interaction between a passing star
and a test star.

The (non-resonant) relaxation time $T_{\mathrm{NR}}$ can be defined as
the time it takes for the energy $\mE$ of a typical star to change by
order unity.  This is also the time it takes for its specific angular
momentum $J$ to change by an amount of order $J_{c}(\mE)$, the maximal
angular momentum for that energy. On Keplerian orbits
$J_{c}\!=\!\sqrt{G\Mbh a}$, where $a$ is the semi-major axis. The
{}``non-resonant'' relaxation time $T_{\mathrm{NR}}$ of stars of mass
$\Ms$ can be written in the Keplerian regime as

\begin{equation}
T_{\mathrm{NR}}=A_{\Lambda}\left({\frac{\Mbh}{\Ms}}\right)^{2}{\frac{P(a)}{N(<a)}}\qquad(\Mbh\!\gg\!\Ms)\,,\label{e:tr}\end{equation}
 where $P\!=\!2\pi\sqrt{a^{3}/(G\Mbh)}$ is the orbital period and
 $A_{\Lambda}$ is a dimensionless constant which includes the Coulomb
 logarithm. For some stellar systems (like galaxies), $T_{\rm NR}$ is
 much longer than the age of the system, implying that the system
 cannot evolve significantly towards steady state by two-body
 interactions. For other systems, like very dense stellar clusters,
 the relaxation time can be as small as a few Myr, and such systems
 may even evaporate within a Hubble time. In our GC, the relaxation
 time is somewhat smaller than the age of the system, $T_{\rm
 NR}\sim{\rm few}\Gyr$ (e.g. Alexander \cite{A99}, \cite{A05}),
 implying that the system has evolved considerably, and two-body
 relaxation effects such as mass-segregation have occurred (Bahcall \&
 Wolf \cite{BW77}; Freitag, Amaro-Seoane \& Kalogera \cite{Fre06};
 Hopman \& Alexander \cite{Hop06b}). At the same time, the relaxation
 time is much longer than some other relevant times in the GC, in
 particular the age of the youngest stars.

The assumption of uncorrelated two-body interactions is well-justified
in many systems, such as globular clusters, where stellar orbits are
not closed. However, the special symmetry of a Keplerian potential
leads to closed, elliptical orbits. The fact that the orbits are
closed can be exploited in numerical treatment, and also leads to
unique dynamical features (see also the contribution of Touma in this
volume). Since $t_r\gg t_d$, the orbits remain closed for many
dynamical times, the system may be thought of as a set of ``wires''
with the mass of the star smeared out over the orbits. In this
picture, it is the wires that interact and cause the evolution of the
system, rather than point particles interacting at given
locations. The idea is reminiscent of the Kozai mechanism in triple
stars (Kozai \cite{KO62}).

Rauch \& Tremaine \cite{RT96} first used this approach in the context
of many body stellar dynamics near MBHs, and coined the term {\it
resonant relaxation} (RR), after the $1\!:\!1$ resonance between the
radial and azimuthal frequencies in a Keplerian potential. The wire
approximation is only relevant for times $\ll \tp$, where $\tp$ is the
time for the orbit to precess. Precession may be caused by the fact
that the potential is not entirely determined by a point mass, and
there is still some extended component to the potential due to the
stellar mass; this is especially the case far away ($\gtrsim0.1\pc$)
from the MBH. Closer to the MBH ($\lesssim0.01\pc$), precession may be
dominated by effects of General Relativity\footnote{For the parameters
of interest here, Lense-Thirring precession is much less efficient
than mass and GR precession even for a maximally spinning MBH, and we
do not consider it here.}.

\subsection{Scalar resonant relaxation}
\label{ss:TRRs}

Scalar relaxation results in changes in \emph{both} the direction and
the magnitude of the angular momenta. The RR time
$T_{\mathrm{RR}}$ is estimated by evaluating $\Delta\Jp$, the coherent
change in the magnitude of the specific angular momentum up to a time
$\tp$. The change $\Delta\Jp$ is then the step size for the
non-coherent growth of the angular momentum over times
$t\!>\!\tp$. Two nearby stars with semi-major axes $a$ exert a mutual
specific torque $\sim\! G\Ms/a$. To zeroth order the torques of the
stars on a test wire cancel, so that within a distance $a$ from the
MBH the net torque on a test star is determined by the Poissonian
excess torque $\dot{J}\!\sim\!\sqrt{N(<\! a)}G\Ms/a$ and

\begin{equation}
\Delta\Jp\sim\dot{J}\tp=\sqrt{N(<\!
a)}(G\Ms/a)\tp\,.\label{e:Jw}
\end{equation} 
For $t\!>\!\tp$ the torques on a particular star-wire become random,
and the change in angular momentum grows in a random walk fashion with
a timescale $T_{\mathrm{RR}}\!\sim\!(J_{c}/\Delta J_{\omega})^{2}\tp$,
defined as

\begin{equation}
T_{\mathrm{RR}}\!\equiv\! A_{\mathrm{RR}}\frac{N(>\!\mE)}{\mu^{2}(>\!\mE)}\frac{P^{2}(\mE)}{\tp}\!\simeq\!\frac{A_{\mathrm{RR}}}{N(<\! a)}\left(\frac{\Mbh}{\Ms}\right)^{2}\frac{P^{2}(a)}{\tp}\,,\label{e:TRR}
\end{equation}
where $\mu\!\equiv\! N\Ms/(\Mbh\!+\! N\Ms)$, $A_{\mathrm{RR}}$ is a
numerical factor of order unity, to be determined by simulations, and
the last approximate equality holds in the Keplerian regime.

Over most of the relevant phase space the precession is due to the
deviations from pure Keplerian motion caused by the potential of the
extended stellar cluster. This occurs on a timescale $\tp\!=\!
t_{M}=[\Mbh/N(<\! a)\Ms]P(a)$, assuming $N(<\! a)\Ms\!\ll\!\Mbh$. The
$J$-averaged RR timescale can then be written as

\begin{equation}
T_{\mathrm{RR}}^{M}=A_{\mathrm{RR}}{\frac{\Mbh}{\Ms}}P(a)={\frac{A_{\mathrm{RR}}}{A_{\Lambda}}}{\frac{\Ms}{\Mbh}}N(<\!
a)T_{\mathrm{NR}}.\label{e:TRRM}\end{equation} Since
$T_{\mathrm{RR}}^{M}\!\ll\! T_{\mathrm{NR}}$ for small $a$ where
$N(<\! a)\Ms\!\ll\!\Mbh$, the RR rate of angular
momentum relaxation is much higher than the rate of energy relaxation
in the resonant regime. This qualitative analysis has been verified by
detailed numerical $N$-body simulations by Rauch \& Tremaine
\cite{RT96} and by Rauch \& Ingalls \cite{RI98}.

For most of parameter space, orbital precession is dominated by the
mass of the stellar cluster and the RR timescale is well approximated
by $T_{\mathrm{RR}}\!\sim\! T_{\mathrm{RR}}^{M}$. However, very close
to the MBH, or on wide orbits with very low angular momentum, so that
the periapse is close to the Schwarzschild radius of the MBH,
precession is dominated by GR effects. In this case the timescale for
precession is given by $\tp\!=\!
t_{\mathrm{GR}}=(8/3)(J/J_{\mathrm{LSO}})^{2}P$; here
$J_{\mathrm{LSO}}\equiv(4G\Mbh/c)$ is the angular momentum of the last
stable orbit (LSO). When $t_{\mathrm{GR}}\!\ll\! t_{M}$ and GR
precession dominates, the RR timescale is (Eq. \ref{e:TRRM})
\begin{equation}
T_{\mathrm{RR}}^{\mathrm{GR}}=\frac{3}{8}A_{\mathrm{RR}}\left(\frac{\Mbh}{\Ms}\right)^{2}\left(\frac{J_{\mathrm{LSO}}}{J}\right)^{2}\frac{P(a)}{N(<\!
a)}\,.\label{e:TRRGR}\end{equation}

Generally, GR precession and mass precession occur simultaneously, and
the scalar RR timescale $T_{\mathrm{RR}}^{s}(\mE,J)$
is given by substituting
$1/\tp\!=\!\left|1/t_{M}-1/t_{\mathrm{GR}}\right|$ in
Eq. (\ref{e:TRR}), where the opposite signs reflect the fact that mass
precession is retrograde whereas GR precession is prograde. Thus, the
scalar RR timescale is\begin{equation}
T_{\mathrm{RR}}^{s}=\frac{A_{\mathrm{RR}}}{N(<\!
a)}\left(\frac{\Mbh}{\Ms}\right)^{2}P^{2}(a)\left|\frac{1}{t_{M}}-\frac{1}{t_{\mathrm{GR}}}\right|\,.\label{e:TRRs}\end{equation}

We use the relation
$\mathrm{d}(J^{2})/J_{c}^{2}\!=\!\mathrm{d}t/T_{\mathrm{RR}}^{s}(\mE,J)$
(Eqs. \ref{e:TRRs}) to define the $J$-averaged time it takes a star to
random-walk from $J=J_{c}(\mE)$ to the loss-cone $J=J_{lc}$ as

\begin{equation}
\bar{T}_{\mathrm{RR}}^{s}(\mE)=\frac{1}{J_{c}^{2}}\int_{J_{lc}^{2}}^{J_{c}^{2}}dJ^{2}T_{\mathrm{RR}}^{s}(\mE,J)\,.\label{e:Tave}\end{equation}

\subsection{Vector resonant relaxation}

\label{ss:TRRv} 

For time scales much larger than the dynamical time, orbits precess
and describe a rosette shape. One can then consider the torques
between different rosettes rather than between different wires. Since
the rosettes describe planar rings to good approximation, they cannot
modify the magnitude of the angular momentum of the star, but they can
change the direction of the angular momentum vector. This process is
known as ``vector resonant relaxation'' (Rauch \& Tremaine
\cite{RT96}).  Vector RR grows coherently ($\propto\! t$) on
timescales $t\!\ll\!  t_{\varphi}$, where $t_{\varphi}$ is the
timescale for a change of order unity in the total gravitational
potential $\varphi$ caused by the changes in the stellar potential
$\varphi_{\star}$ due to the realignment of the stars as they rotate
by $\pi$ on their orbit,\begin{equation}
t_{\varphi}=\frac{\varphi}{\dot{\varphi_{\star}}}\simeq\frac{N^{1/2}}{\mu}\frac{P}{2}\simeq\frac{1}{2}\frac{\Mbh}{\Ms}\frac{P}{N^{1/2}}\,,\label{e:tphi}\end{equation}
the last approximate equality holds for $N\Ms\!\ll\!\Mbh$. In analogy
to scalar RR (Eq. \ref{e:Jw}), the maximal coherent change in
$\mathbf{J}$ is
$\left|\Delta\mathbf{J}_{\varphi}\right|\sim\dot{J}t_{\varphi}\sim
J_{c}$, that is, $\mathbf{J}$ rotates by an angle ${\cal {O}}(1)$
already at the coherent phase. On timescales $t\!\gg\!  t_{\varphi}$,
$\left|\Delta\mathbf{J}_{\varphi}\right|$ cannot grow larger, as it
already reached its maximal possible value, but the orbital
inclination angle is continuously randomized non-coherently
($\propto\! t^{1/2}$) on the vector RR timescale (Eq. \ref{e:TRR}),

\begin{equation}
T_{\mathrm{RR}}^{v}=2A_{\mathrm{RR}}^{v}\frac{N^{1/2}(>\!\mE)}{\mu(\mE)}P(\mE)\simeq2A_{\mathrm{RR}}^{v}\left(\frac{\Mbh}{\Ms}\right)\frac{P(a)}{N^{1/2}(<\! a)},\label{e:Tv}\end{equation}
 where the last approximate equality holds for $N\Ms\!\ll\!\Mbh$. 

It is that while the torques driving scalar and vector resonant
relaxation are the same, vector RR is much more
efficient than scalar RR,
$T_{\mathrm{RR}}^{v}\!\ll\!\bar{T}_{\mathrm{RR}}^{s}$, due to the much
longer coherence time $t_{\varphi}\!\sim\! N^{1/2}t_{M}\!\gg\! t_{M}$.
Furthermore, vector RR proceeds irrespective of any
precession mechanisms that limit the efficiency of scalar resonant
relaxation.

\section{The origin of the young stellar population in the Galactic center}

\begin{figure}[h]
\includegraphics[width=18pc]{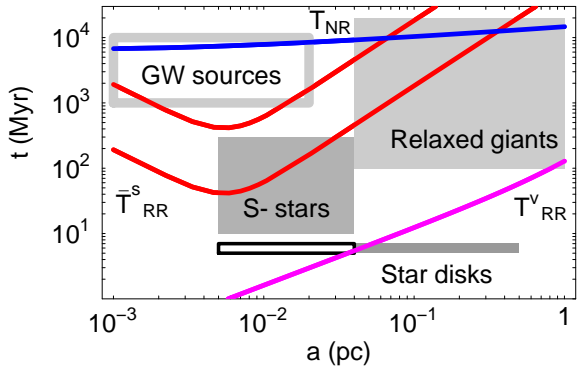}
\hspace{2pc}
\begin{minipage}[b]{18pc}
\caption{\label{f:GC}Stellar components, timescales and distance
scales in the GC. The NR timescale $T_{\mathrm{NR}}$ (top straight
line); the timescale $\bar{T}_{\textrm{RR}}^{s}$, estimated for
$1\,\Mo$ stars (top curved line) and $10\,\Mo$ stars (bottom curved
line); the timescale $T_{\mathrm{RR}}^{v}$ (bottom straight line); the
position and estimated age of the young stellar rings in the GC
(filled rectangle in the bottom right); the position and age of the
S-stars if they were born with the disks (empty rectangle in the
bottom left); the position and maximal lifespan of the S-stars (filled
rectangle in the middle left). Reprinted with permission from the
Astrophysical Journal}
\end{minipage}
\end{figure}

Figure (\ref{f:GC}) compares the distance scales and the ages or
lifespans of the various dynamical structures and components in the
inner pc of the GC with the relaxation timescales. The NR timescale in
the GC, which is roughly independent of radius, is
$T_{\mathrm{NR}}\!\sim\!\mathrm{few\!\times\!10^{9}}\,\mathrm{yr}$
(Eq. \ref{e:tr}). The scalar RR time $\bar{T}_{\mathrm{RR}}^{s}$ is
shown for $\Ms=1,10\,\Mo$. At large radii the RR time decreases
towards the center, but for small radii, where GR precession becomes
significant, it increases again. The vector RR timescale
$T_{\mathrm{RR}}^{v}$, in contrast, decreases unquenched with
decreasing radius. Structures with estimated ages exceeding these
relaxation timescales must be relaxed.

Two distinct young stellar populations exists in the GC. At distances
of $0.04$--$0.5$ pc from the MBH there are about $\sim\!70$ young
massive OB stars ($\Ms\!\gg\!10\,\Mo$, lifespan of
$t_{\star}\!=\!6\pm2$ Myr), which are distributed in two nearly
perpendicular, tangentially rotating disks (Levin \& Belobodorov
\cite{LB03}; Genzel et al. \cite{Gea03}; Paumard et al. \cite{P05}).
It appears that these stars were formed by the fragmentation of gas
disks (Levin \& Belobodorov \cite{LB03}; Levin \cite{L03}; Nayakshin
\& Cuadra \cite{NC04}; Nayakshin \& Sunyaev \cite{NS05}; Nayakshin
\cite{N06}). Inside the inner $0.04\,\mathrm{pc}$ the population
changes. There is no evidence for old stars, and the young stars there
(the {}``S-stars'') are main-sequence B-stars ($\Ms\!\lesssim15\,\Mo$,
lifespans of $10^{7}\mathrm{\!\lesssim
t_{\star}\!\lesssim\!2}\!\times\!10^{8}$ yr; Ghez et al. \cite{Gh03};
Eisenhauer et al. \cite{Ei05}) on randomly oriented orbits with a
random (thermal) $J$-distribution. There is to date no satisfactory
explanation for the presence of the S-stars so close to the MBH (see
Alexander \cite{A05} for a review). 

The existence of coherent dynamical structures in the GC constrains
the relaxation processes on these distance scales, since the the
relaxation timescales must be longer than the structure age
$t_{\star}$ to avoid randomizing it. Figure (\ref{f:GC}) shows that
the observed systematic trends in the spatial distribution, age and
state of relaxation of the different stellar components of the GC are
consistent with, and perhaps even caused by RR. The star disks are
young enough to retain their structure up to their inner edge at
$0.04\,\mathrm{pc}$, where $t_{\star}\!\sim\!  T_{\mathrm{RR}}^{v}$
and vector RR can randomize the disk (Hopman \& Alexander
\cite{Hop06a}). It is tempting to explain the S-stars as originally
being the inner part of the same disks that are currently present in
the GC. However, this scenario is somewhat problematic. First, we note
that vector relaxation can only change the inclinations of the orbits,
and not their eccentricities, while many of the S-stars have high
($e>0.9$) eccentricities; the scalar resonant relaxation time is
larger than the age of the disks. Second, resonant relaxation alone
cannot explain why the S-stars are systematically less massive than
the disk stars. An alternative (Levin \cite{L06}) would be that the
S-stars were perhaps formed in previous accretion disks of which the
dynamical signatures have now disappeared.

If the S-stars were {\it not} formed in the disk, but captured by
either a tidal binary disruption (Gould \& Quillen \cite{GQ03}; see
also contribution from Perets et al. in this volume) or an exchange
interaction with a stellar mass black hole (Alexander \& Livio
\cite{AL04}), they may be much older than the disks, and in particular
their age may be comparable to the local scalar RR time (see figure
\ref{f:GC}). In this case, RR will redistribute their orbits within
their life-time. This may be an essential element of these formation
mechanisms: both scenarios lead to rather eccentric orbits (especially
tidal binary disruption), whereas not all the orbits of the S-stars
are very eccentric: star S1 has eccentricity $e=0.358\pm0.036$, and
S31 has $e=0.395\pm0.032$ (Eisenhauer et al. \cite{Ei05}). Since the
age of these stars may well exceed the RR time, RR may have
redistributed the eccentricities to the current DF, which is
consistent with a thermal DF.

Regardless of the origin of the S-stars, their random orbits are
consistent with the effect of RR. Vector RR can also explain why the
evolved red giants beyond $~\!0.04\,\mathrm{pc}$, in particular the
more massive ones with
$t_{\star}\!\ll\!\min(T_{\mathrm{NR}},\bar{T}_{\mathrm{RR}}^{s})$ are
relaxed, since $T_{\mathrm{RR}}^{v}\!<\! t_{\star}$ out to $\sim\!1$
pc.

\section{Gravitational wave sources}

MBHs with masses $\Mbh\lesssim5\times10^6\Mo$ have Schwarzschild radii
$r_S=2G\Mbh/c^2$, such that a test mass orbiting at a few $r_S$ emits
gravitational waves (GWs) with frequencies $10^{-4}{\rm
Hz}\!\lesssim\!\nu\!\lesssim\!1{\rm Hz}$, detectable by the planned
space based {\it Laser Interferometer Space
Antenna}\footnote{http://lisa.jpl.nasa.gov/} ({\it LISA}). Such GW
sources, for which the mass of the inspiraling object is many orders
of magnitude smaller than the mass of the MBH are known as {\it
extreme mass ratio inspiral sources} (EMRIs). GW inspiral events are
very rare (of the order of $10^{-7}-10^{-8}{\rm \,yr^{-1}}$ per
galactic nucleus; e.g. Hils \& Bender \cite{HB95}; Sigurdsson \& Rees
\cite{SR97}; Ivanov \cite{IV02}; Freitag \cite{FR01}; Alexander \&
Hopman \cite{AH03}; Hopman \& Alexander \cite{HA05}, \cite{Hop06a},
\cite{Hop06b}), and it is unlikely that we will observe GWs from our
own Galactic center (GC), although such a possibility is not entirely
excluded (Freitag \cite{FR03}; Rubbo, Holley-Bockelmann \& Finn
\cite{Rub06}). The galactic MBH plays nevertheless a role of
importance in understanding the dynamics of EMRIs, since its mass is
very close to the mass of the ``optimal'' {\it LISA} EMRI target, and
as a consequence one may use the GC to model extra-galactic nuclei.

Hopman \& Alexander \cite{HA05} used a model based on the GC to
analyze the dynamics of EMRIs. One of the main results was that
inspiraling stars always originate very near the MBH, within a
distance of $\sim0.01\pc$: due to the relatively short relaxation time
in galactic nuclei, stars that start to spiral in from larger
distances are very likely to plunge into the MBH before becoming
observable as GW emitters. This result was confirmed qualitatively by
$N$-body simulations by Baumgardt et al. \cite{Bea05} of tidal capture
of MS stars by an intermediate mass black hole (Hopman, Portegies
Zwart \& Alexander \cite{HPZA04}; Hopman \& Portegies Zwart
\cite{HPZ05}).

The fact that only stars within $\sim0.01\pc$ spiral in successfully,
implies that it is the stellar content and dynamics of that region
which determine the rate of GW inspiral events for the different
populations in the system. This means, for example, that
mass-segregation is likely to play an important role (Hopman \&
Alexander \cite{Hop06b}; Freitag et al.; \cite{Fre06}; see also
contribution by Marc Freitag in this volume). Since the resonant
relaxation time is very short ($T_{\rm RR}\ll T_{\rm NR}$) near
$\sim0.01\pc$, it also implies that RR will dictate the rate at which
stars are driven towards low $J$ orbits, where energy dissipation is
efficient and stars spiral in.

Hopman \& Alexander \cite{Hop06a} used a Fokker-Planck method in
energy space with a sink term due to RR losses in $J$-space to
calculate the GW inspiral rate. At every time-step, stars redistribute
in energy-space due to (non-resonant) two body scattering, and stars
are accreted by the MBH with some specified rate per energy bin. In
the relevant regime, this rate is assumed to be of order $\sim
N(E)/T_{\rm RR}$, i.e., within one RR time all stars in the bin would
be accreted if they were not replaced by new stars that flow to higher
energies (tighter orbits). In spite of the fact that stars are drained
very efficiently near the MBH, Hopman \& Alexander \cite{Hop06a} found
that the rate at which stars are replenished by two-body scattering is
sufficiently high that the stellar distribution will not be depleted
near the MBH, unless the efficiency of RR is more than an order of
magnitude larger than exploratory $N$-body simulations (Rauch \&
Tremaine \cite{RT96}) have indicated. Modifications of the stellar DF
due to RR are too small to be observable. Some of the stars that are
captured by the MBH are swallowed directly without giving a GW signal,
but the stars closed to the MBH (within $\sim0.01\pc$) will spiral in
rapidly enough to give obtain an orbit of period $P\lesssim10^4\,{\rm
s}$ for more than a year. Such sources would be observable to {\it
LISA} to distances up to a few Gpc, depending on the mass of the
inspiraling star. Since the enhanced rate at which stars flow to the
loss-cone in angular momentum space is sustained by the larger flow
due to two-body scattering in energy space, the rate at which EMRIs
are produced is increased. The analysis by Hopman \& Alexander
\cite{Hop06a} indicates that the rate at which {\it observable} EMRIs
are formed in galactic nuclei is $\sim8$ times higher than that for
the case in which RR was neglected.

\section{Conclusions}

Resonant relaxation is a relatively unexplored dynamical mechanism.
Vector RR, which only affects the orientation of the orbit but not the
eccentricity, operates in many stellar systems, while scalar RR, which
does affect the eccentricity, is unique for MBH systems. This
mechanism becomes important at distances $\lesssim0.1\pc$ from the
MBH. It may have played an important role in redistributing the orbits
of the S-stars, and enhances estimates of the GW inspiral rate by
nearly an order of magnitude. Our own GC provides a unique case study
for resonant relaxation.

\ack We thank the organizers of the GC2006 meeting for a very
stimulating conference, and Yuri Levin for discussions on resonant
relaxation.

\end{document}